\documentclass{aastex62}
\pdfoutput=1
\usepackage{amsmath}
\usepackage{xcolor}

\newcommand{\nuovernuref}{{\left(\frac{\nu}{\nu_{\text{ref}}}\right)}}

\graphicspath{{./}{figures/}}

\received{ xxx 2018}
\revised{xxx, 2018}
\accepted{xxx, 2018}
\submitjournal{ApJ}

\shorttitle{Ground Plane Artifact in Global 21-cm Measurements}
\shortauthors{Bradley et al.}

\begin{document}

\title{A Ground Plane Artifact that Induces an Absorption Profile in Averaged Spectra from Global 21-cm Measurements - with Possible Application to EDGES}

\correspondingauthor{Richard F. Bradley}
\email{rbradley@nrao.edu, rfb7m@virginia.edu}

\author[0000-0002-0786-7307]{Richard F. Bradley}
\affiliation{National Radio Astronomy Observatory, NRAO Technology Center, 1180 Boxwood Estate Road, Charlottesville, VA 22903-4602, USA}

\author[0000-0003-1932-9829]{Keith Tauscher}
\affiliation{Center for Astrophysics and Space Astronomy, Department of Astrophysical and Planetary Sciences, University of Colorado, Boulder, CO 80309, USA}
\affiliation{Department of Physics, University of Colorado, Boulder, CO 80309, USA}

\author[0000-0003-2196-6675]{David Rapetti}
\affiliation{Center for Astrophysics and Space Astronomy, Department of Astrophysical and Planetary Sciences, University of Colorado, Boulder, CO 80309, USA}
\affiliation{NASA Ames Research Center, Moffett Field, CA 94035, USA}

\author[0000-0002-4468-2117]{Jack O. Burns}
\affiliation{Center for Astrophysics and Space Astronomy, Department of Astrophysical and Planetary Sciences, University of Colorado, Boulder, CO 80309, USA}



\begin{abstract}
Most of the current Global 21-cm experiments include ground screens that help moderate effects from the Earth.  In this paper, we report on a possible systematic artifact within the ground plane that may produce broad absorption features in the spectra observed by these experiments.  Using analytical approximations and numerical modeling, the origin of the artifact and its impact on the sky-averaged spectrum are described. The publicly released EDGES dataset, from which a 78 MHz absorption feature was recently suggested, is used to probe for the potential presence of ground plane resonances. While the lack of a noise level for the EDGES spectrum makes traditional goodness-of-fit statistics unattainable, the rms residual can be used to assess the relative goodness of fits performed under similar circumstances. The fit to the EDGES spectrum using a model with a simple 2-term foreground and three cavity-mode resonances is compared to a fit to the same spectrum with a model used by the EDGES team consisting of a 5-term foreground and a flattened Gaussian signal. The fits with the physically motivated resonance and empirical flattened Gaussian models have rms residuals of 20.8 mK (11 parameters) and 24.5 mK (9 parameters), respectively, allowing us to conclude that ground plane resonances constitute another plausible explanation for the EDGES data.
\end{abstract}

\keywords{instrumentation: detectors --- cosmology: observations}


\section{Introduction} \label{sec:intro}
The weak nature of the highly redshifted 21-cm\ signature in the presence of strong foreground radiation, instrument systematics, and radio frequency interference makes the detection of a spectral feature in this band challenging in terms of measurement sensitivity and calibration. The EDGES (Experiment to Detect the Global Epoch of Reionization Signature) collaboration, in \cite{bowman2018absorption}, recently published a sky-averaged spectrum covering 50-100 MHz, revealing an absorption feature near 78 MHz ($z \approx 17$) that may have important cosmological implications if its astrophysical origin can be verified \citep{Fraser:2018,Kovetz:2018,Ewall-Wice:2018}. However, very subtle systematic errors within the instrument can have a profound effect on the interpretation of the data, if not fully encapsulated by the calibration.  For example, \cite{hills2018concerns} described a possible fit to the released EDGES data without any absorption trough when there is a periodic feature with an amplitude of $\approx 0.05$ K in the sky spectrum. Here, we attempt to address the question: Is there a systematic artifact within the EDGES instrument that could produce such an absorption trough?

We explored the viability of a ground plane artifact producing an energy absorbing feature at 78 MHz. We first analyzed the geometrical and material properties of the ground plane and soil from an electromagnetic perspective, leading to a lossy resonant cavity model, i.e. a patch antenna, with specific modes that exhibit spectral absorption features at frequencies defined by physical parameters.   While parameters related to the ground plane's horizontal dimensions are well defined, those involving the soil properties (dielectric permittivity and loss) and depth are less so, instead being bounded within a range of values estimated from measured soil samples in the vicinity of the instrument.   Nonetheless, this analysis motivated an independent search for resonances within the EDGES data, where we found that three such features along with a two-term foreground model fit the data with rms residuals similar in magnitude to that of the five-term foreground and flattened Gaussian signal model postulated by the EDGES team.  Without the noise level of the data and thus the ability to compute a goodness-of-fit statistic \citep[such as the traditional $\chi^2$, or $\psi^2$ as defined in][]{Tauscher:2018b}, however, the significance of the differences in parameter count and rms residual level between the two fits can not be determined. We extracted the cavity parameters from the resonant features fit to the EDGES data, yielding values that agree with the EDGES central solid ground plane horizontal dimensions and lie within the expected ranges for the soil characteristics.


In this paper, the physical characteristics of the artifact are described in Sec.~\ref{sec:Patch}, and the model-based analysis is presented in Sec.~\ref{sec:Antenna}. Details of the procedure and results of the  independent, least-squares fit of three resonance features to the published EDGES data are provided in Sec.~\ref{sec:Fit} along with a comparison to a similar fit for a flattened-Gaussian model. Estimations for the physical resonance model parameters are calculated in Sec.~\ref{sec:Physical}. Further considerations on initial experimental and simulation tests are discussed in Sec.~\ref{sec:Consider}, followed by concluding remarks in Sec.~\ref{sec:Conclusion}.

\section{Physical Characteristics of the Ground Plane Artifact}
\label{sec:Patch}
\begin{figure}[b!]
\centering
\includegraphics[width=\textwidth]{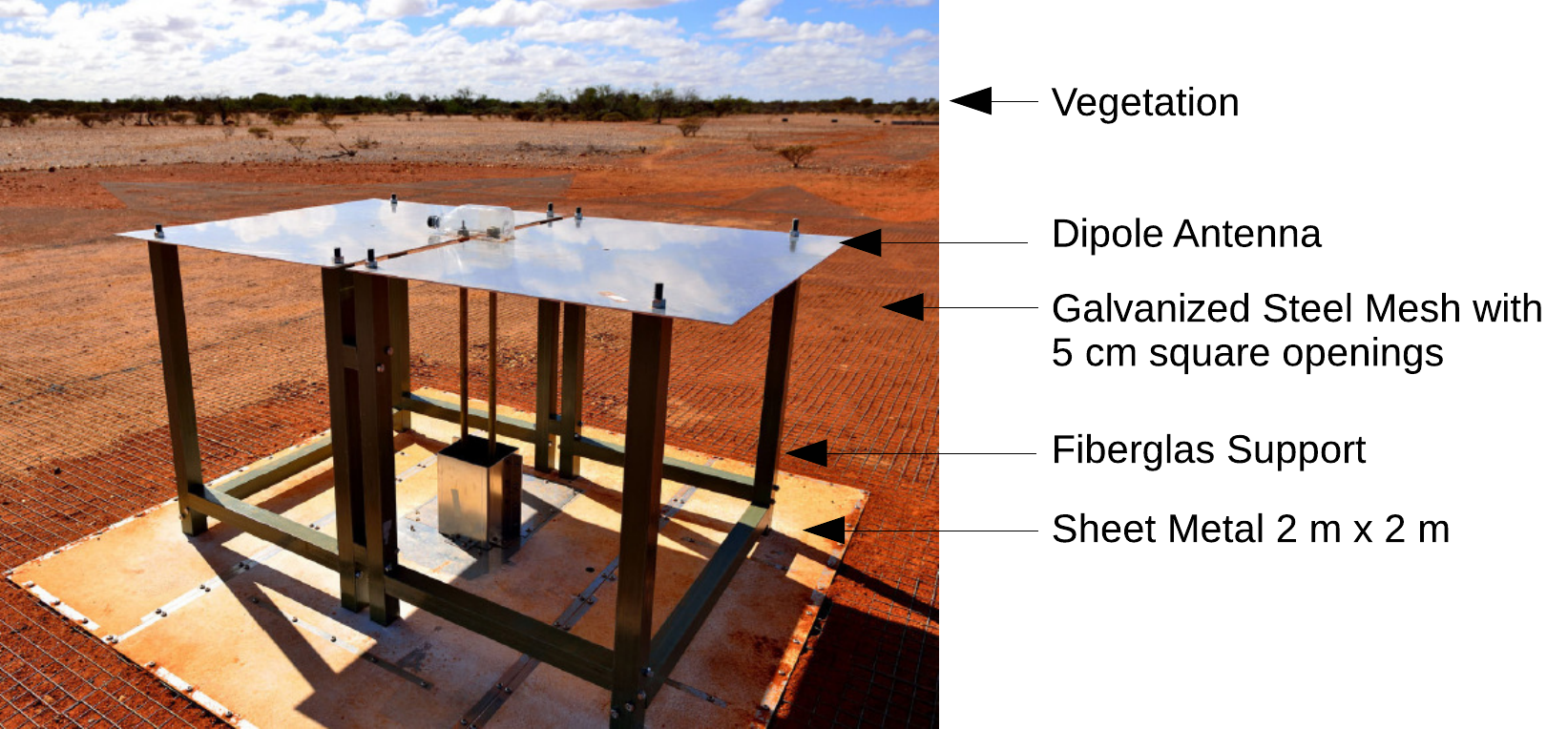}
\caption{Photograph of the EDGES antenna system with the major components labeled. The mesh extends to 30 $m$ $\times$ 30 $m$, with the outer 5 $m$ shaped as saw-tooth perforated edges.}
\label{fig:EDGESphoto}
\end{figure}

A photograph of the EDGES instrument is shown in Fig. \ref{fig:EDGESphoto}. 
The antenna system consists of a pair of horizontally-oriented, flat metal plates forming the dipole element suspended above the ground plane by a Fiberglas supporting structure. The ground plane consists of two components: a 2 $m$ square inner portion of solid sheet metal and a larger mesh that extends out to approximately 30 $m$. The entire antenna and ground system rest directly on the soil having characteristics summarized in Table \ref{tab:soil}. These were based on measured soil samples from the Murchison Radioastronomy Observatory (MRO) that were characterized for several moisture levels at spot frequencies \citep{sutinjo2015characterization}.  Moist soil must exist in the vicinity of the instrument to support the vegetation present in Fig. \ref{fig:EDGESphoto}.

\begin{table}[h!]
\renewcommand{\thetable}{\arabic{table}}
\centering
\caption{Measured Relative Permittivity $\epsilon_r$ and Conductivity $\sigma$ (S/m) of MRO Soil for Several Moisture Levels.}
\label{tab:soil}
\begin{tabular}{ccccccc}
\tablewidth{0pt}
\hline
\hline
MHz & \multicolumn{2}{c}{Dry} & \multicolumn{2}{c}{1 $\%$} & \multicolumn{2}{c}{$10 \%$} \\
& $\epsilon_r$ & $\sigma$ & $\epsilon_r$ & $\sigma$ & $\epsilon_r$ & $\sigma$ \\
\hline
\decimals
50 & 3.9 & 0.0007 & 6.5 & 0.01 & 17.8 & 0.1  \\
160 & 3.7 & 0.0018 & 5.2 & 0.017 & 14.8 & 0.11 \\
280 & 3.7 & 0.0022 & 4.8 & 0.02 & 14.4 & 0.11 \\
\hline
\multicolumn{7}{c}{From \cite{sutinjo2015characterization}}
\end{tabular}
\end{table}

\begin{figure}[h!]
\centering
\includegraphics[width=0.8\textwidth]{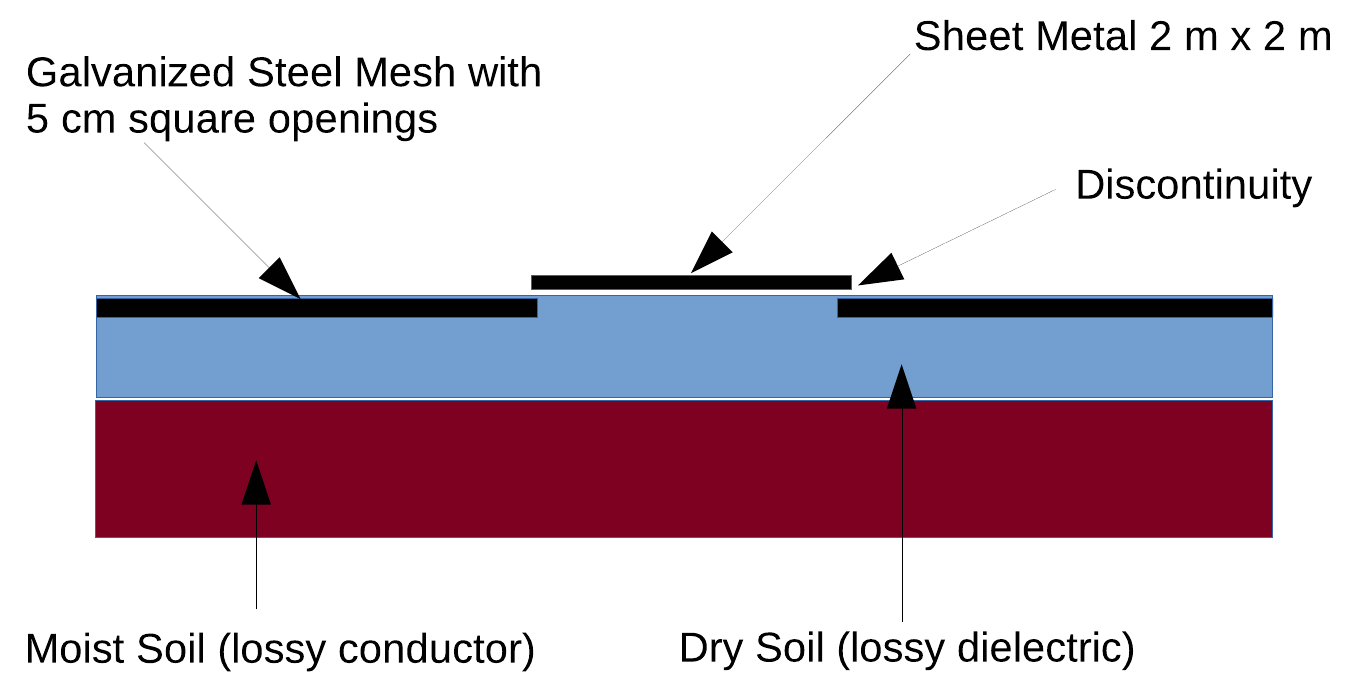}
\caption{A sketch of the ground plane region of the EDGES instrument showing the soil layers and ground discontinuity between the sheet metal and galvanized, welded steel mesh.}
\label{fig:EDGESsketch}
\end{figure}

The artifact secluded within the instrument is illustrated by the cross-sectional sketch of the ground plane and soil shown in Fig. \ref{fig:EDGESsketch}. The metal ground plane (solid sheet and welded steel mesh) and subsurface moist soil form the upper and lower conductors of a lossy transmission line, with the dry soil between these two conductors as the dielectric material. While waves may propagate under the entire ground plane via the transverse electromagnetic (TEM) mode that exists on this transmission line, it is the central sheet metal region that is of particular interest.

Careful inspection of Fig. \ref{fig:EDGESphoto} shows the aluminum sheet metal portion of the ground plane lies slightly above the galvanized welded steel mesh.  Since the soil near this interface is not compacted, persons walking near the instrument to service the electronics will force the mesh partway into the soil and can easily loosen or break the connections between the two ground structures.  Galvanic interaction between the two dissimilar metals may also be a factor. For example, $ZnO$, formed at the interface by this process, is a wide-bandgap (3.35 $eV$), {\it II-VI} group semiconductor with native $n$-type doping that could form a Schottky barrier that effectively blocks weak signals \citep{szephysics}.   Unfortunately, due to its design, the integrity of this interface cannot be inspected from above.

It should be noted that soil moisture is not the only possibility for the lower conductor.  For example, a thin seam of a metallic mineral can also serve this purpose.  A careful analysis of the soil structure under ground planes using trenching or ground penetrating radar techniques is required for an accurate assessment.  Alternatively, a synthetic ground plane foundation should be considered to carefully engineer subsurface electromagnetic properties.  

If an electrical discontinuity does exist between the sheet metal and mesh components of the ground plane, a resonant cavity or short section of transmission line is formed that supports a TEM-like mode of propagation \citep{pozar1990microwave}.  Since this line is open at the discontinuities, the structure becomes a harmonic oscillator, resonant at select frequencies that are multiples of half the guide wavelength within the material.  Fringing electric fields at the edges of the sheet provide access to the upper half-space above the ground plane, where the resonator can be excited by the impinging waves.  Hence, this artifact, known as a resonant patch antenna, extracts and dissipates a small fraction of the sky radiation from the waves that reach the ground plane, i.e. {\it energy is absorbed from the reflected waves in a frequency and spatially dependent manner, producing an absorption profile in the dipole-measured spectrum that is characterized by the geometry and material properties of this resonator}.    

\section{Analytical Formulation of the Patch Antenna }
\label{sec:Antenna}

The rectangular patch antenna may be analyzed as two orthogonal transmission lines that are open at both ends to form a two-dimensional resonant structure.  The central region is square for EDGES and will respond equally to the two linear polarizations of the impinging sky radiation. For this analysis, a Cartesian coordinate system is imposed having its origin at the corner of the square and oriented such that the $x$ and $y$ axes align to the ground plane sides with $z$ as the vertical axis.  Transverse Magnetic (TM) waves, having a $z$-directed E-field, can propagate along either the $x$ or $y$ transmission line separately (TM10, TM01, TM20, etc.) or along both simultaneously (TM11, TM22, etc.).  Resonance occurs when the electrical length of the transmission pathways equal multiples of a half-wavelength.

The resonant frequencies, $\nu_{mn}$, of the patch depend upon its geometry and material properties \citep{balanis1997antenna}.  For a lossless (non-dispersive) patch, 
\begin{equation}
\label{eq:resfreq}
\nu_{mn} = \frac{c}{2\pi\sqrt{\epsilon_{re}}}\sqrt{\left( \frac{m\pi}{D} \right)^2 + {\left( \frac{n\pi}{W}\right)^2}}
\end{equation}
where $\epsilon_{re}$ is the effective relative permittivity of the substrate material including the effects of electric field fringing, $c$ is the speed of light {\it in vacuo}, and $m,n$ are positive integers indicating the number of half-cycles variations of the field along the $x$ and $y$ directions. $D$ and $W$ are the horizontal dimensions of the patch.  From \cite{balanis1997antenna}, the fringing field modifies the dielectric constant as
\begin{equation}
\epsilon_{re} = \left( \frac{\epsilon_r+1}{2}\right) + \left(  \frac{\epsilon_r - 1}{2} \right)\left[1 + 12\left(\frac{h}{W}\right)\right]^{-0.5}
\end{equation}
where $\epsilon_r$ is the relative dielectric constant of the dry soil and $h$ is the vertical distance between the ground plane and the underground lower conductor (moist soil). 

The initial estimates of the potential resonant frequencies for the EDGES experiment were calculating assuming a $D=W=2.1$ m square patch (2 m of physical length with 10 cm extra to account for fringing) on top of 40 cm of soil having a dielectric constant $\epsilon_r=4.5$ (yielding $\epsilon_{re} = 3.7$). Not included in this basic analysis is  dispersion resulting from the lossy dielectric and lower conductor.  However, it is included in the analysis of Sec. \ref{sec:Physical} to improve the accuracy of the parameter estimation.  Table \ref{tab:resonances} lists the first nine $TM_{mn}$ resonant frequencies.
\begin{table}[h!]
\renewcommand{\thetable}{\arabic{table}}
\centering
\caption{Initial Estimates of the Resonant Frequencies for the Ground Plane Artifact}
\label{tab:resonances}
\begin{tabular}{cccccc}
\tablewidth{0pt}
\hline
\hline
Mode & m & n & k & $\nu_{mn}$ & Note\\
& & & [m$^{-1}$] & [MHz] & \\
\hline
\decimals
TM10 & 1 & 0 & 1.50 & 37.0 & Fundamental\\
TM11 & 1 & 1 & 2.12 & 52.4 & In-band\\
TM20 & 2 & 0 & 2.99 & 74.1 & In-band\\
TM21 & 2 & 1 & 3.35 & 82.8 & In-band\\
TM22 & 2 & 2 & 4.23 & 105 & Partially In-Band\\
TM30 & 3 & 0 & 4.49 & 111 & Partially In-Band\\
TM31 & 3 & 1 & 4.73 & 117 & \\
TM32 & 3 & 2 & 5.39 & 134 & \\
TM33 & 3 & 3 & 6.35 & 157 & \\
\hline
\end{tabular}
\end{table}

\begin{figure}[t!!]
\centering
\includegraphics[width=0.55\textwidth]{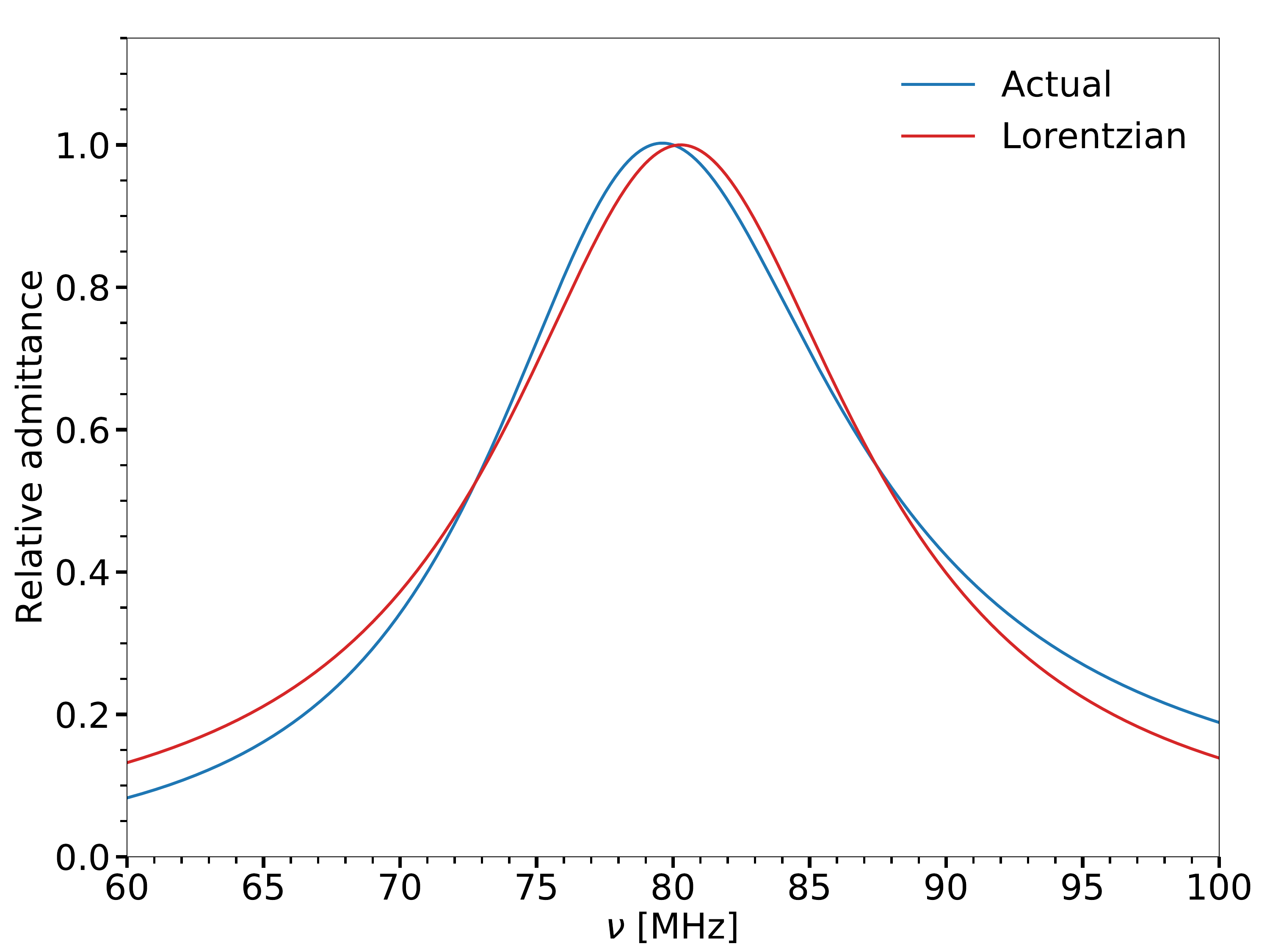}
\caption{Example harmonic oscillator resonance located at $\nu_0=80$ MHz and its least square Lorentzian approximation. The difference between the two becomes significant for low values of $Q$.  Here, $Q=5$.}
\label{fig:lorentzian}
\end{figure}

The spectral shape of each resonance is that of a harmonic oscillator, as shown in Fig. \ref{fig:lorentzian}.
An absorption resonance is characterized by three parameters: $\nu_0$, frequency at which maximum absorption occurs, the depth of the profile $A(\nu_0)$, and Quality Factor, $Q$, which is the ratio of the frequency of maximum absorption to the spectral width of the absorption, usually specified as the Full-Width at Half-Maximum (FWHM).  In low-loss situations ($Q\gg 100$), the Lorentzian curve is traditionally used to approximate the spectral profile $A(\nu)$.  However, in the case of a very lossy resonator such as soil, this approximation is inaccurate and the actual curve

\begin{equation}
\label{eq:reson}
A(\nu) = 1 - \frac{\nu^3\nu_0}{\nu^4+Q^2(\nu^2-\nu_0^2)^2}
\end{equation} 

must be used.

\section{Fitting to EDGES Spectral Data} \label{sec:Fit}
\subsection{Likelihood and minimization procedure}
\begin{figure}[h!!]
    \centering
    \includegraphics[width=0.5\textwidth]{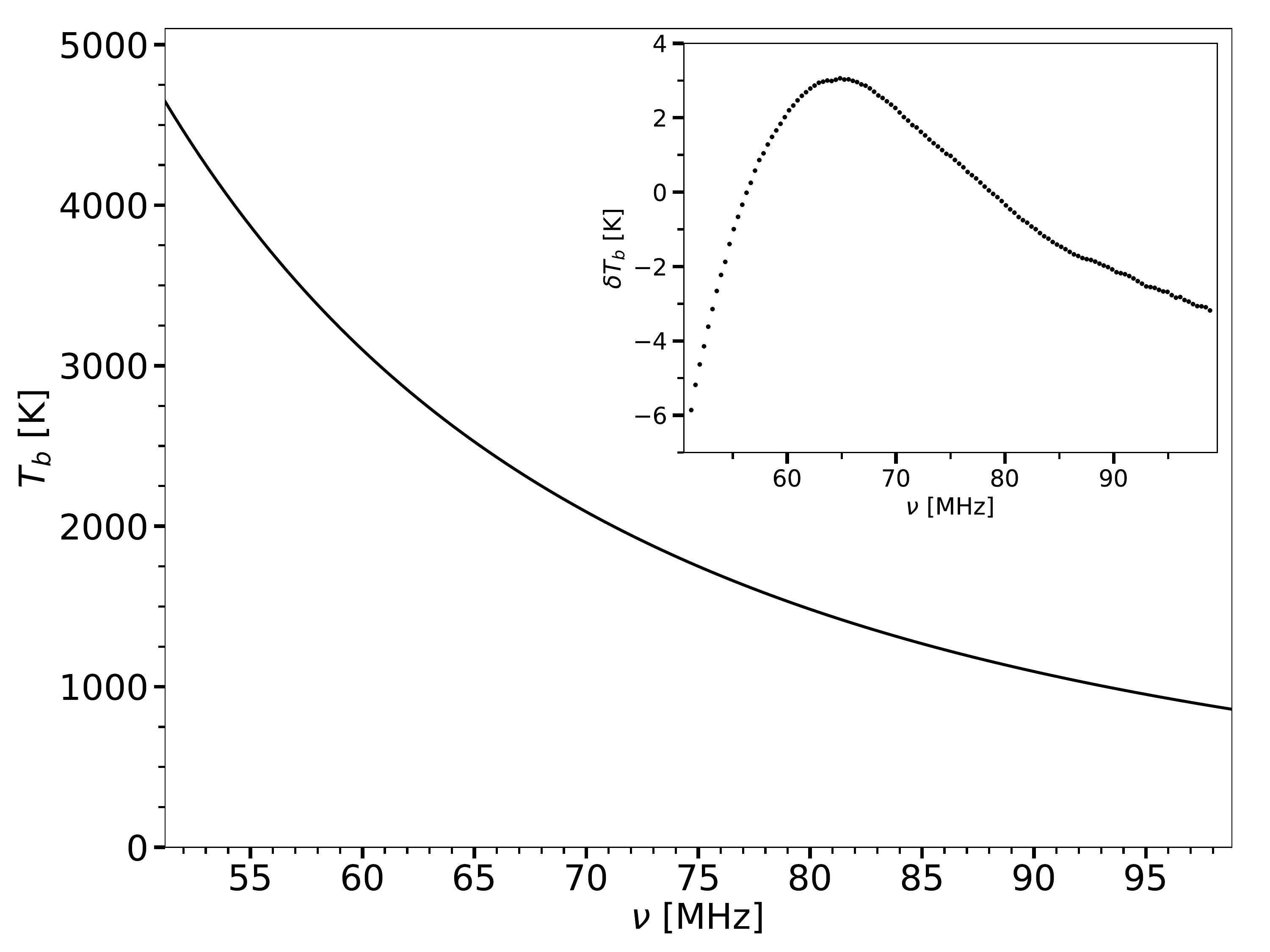}
    \caption{Time averaged brightness temperature spectrum released by \cite{bowman2018absorption} and used for the fitting in this paper. The inset shows the spectrum with the (unweighted) least square fit power law of the form $T(\nu)=T_0\ (\nu/\nu_0)^\alpha$ removed. The spectral index of this fit is $\alpha=-2.560$ and the fit amplitude is $T_0=1749$ K when $\nu_0=75$ MHz.} \label{fig:edges-released-data}
  \end{figure}

To test the EDGES data for the presence of resonances like those described in the previous section, we use gradient descent to fit the data by minimizing a negative log-likelihood which, up to an additive constant, is given by
  \begin{equation}
    -\ln{L} \equiv \frac{1}{2}[{\bf y}-{\bf \mathcal{M}({\bf\theta})}]^T{\bf C}^{-1}[{\bf y}-{\bf \mathcal{M}({\bf\theta})}], \label{eq:fitting-negative-loglikelihood}
  \end{equation}
  where ${\bf y}$ is the spectrum of data released by \cite{bowman2018absorption}, shown in Fig.~\ref{fig:edges-released-data}, ${\bf C}$ is the noise covariance, and ${\bf\mathcal{M}}({\bf\theta})$ is the model of the spectrum at parameters ${\bf\theta}$. In accordance with the radiometer equation, we assume that the standard deviation of the noise is proportional to the data itself \citep[see, e.g.,][]{condon-essential-radio-astronomy}, i.e.
  \begin{equation}
    C_{kl} \propto {y_k}^2\delta_{kl}. \label{eq:channel-covariance}
  \end{equation}
  The precise proportionality factor is unknown as no estimate of the noise level or total integration time was released with the averaged spectrum. But, our choice of proportionality constant is arbitrary because changing the proportionality constant would change only the magnitude of the log-likelihood, not its shape as a function of the parameters, thus leaving maximum likelihood parameter estimation unaffected.
  
  The model ${\bf\mathcal{M}}({\bf\theta})$ is a combination of a concise foreground model and models for the resonances described in this paper. For the foreground, we use a similar model form as \cite{bowman2018absorption}, namely
  \begin{equation}
    T_{\text{fg}} = \nuovernuref^{-2.5}\sum_{k=1}^Na_k \left[\ln{\nuovernuref}\right]^{k-1}. \label{eq:fitting-foreground-model}
  \end{equation}
  The difference between our model and the one used throughout most of \cite{bowman2018absorption} is that we use only $N=2$ terms and the polynomial part is in $\ln{\nu}$ space. {\it We stress that, as more terms are added, polynomial models become increasingly flexible and possibly unphysical, especially in the presence of other models (e.g. 21-cm signal models)}.
  
  The model chosen to represent the possible ground plane resonances in the time-averaged data is given by three copies of the negative term in Eq.~\ref{eq:reson}. The amplitudes, center frequencies, and Q-factors of all three resonances are allowed to vary independently to account for their different beam patterns and the difference in the skin depth of the soil at different frequencies.
  
  We implement model gradients analytically, allowing us to also compute the log-likelihood gradient analytically. We take advantage of this gradient by employing the \texttt{minimize} function of the \texttt{optimize} module of the \texttt{scipy} code\footnote{\url{https://scipy.org}} to perform a gradient descent minimization of the negative loglikelihood given in Eq.~\ref{eq:fitting-negative-loglikelihood}. Because the results of gradient descent procedures depend on the initial guess given to the algorithm, we perform it many times from many different initial parameter vectors and choose from the set of results the final parameter vector which ends up with the largest likelihood value. This is similar to the ``basin hopping'' minimization method of \cite{wales1997basinhopping} in its two-tiered nature, but different in that the individual gradient descent iterations are completely independent of each other. All iterations of the minimization procedure converged on the same region of parameter space, indicating that there is a single dominant global minimum. The fitting method employed here is available for use as part of the \texttt{pylinex} code,\footnote{\url{https://bitbucket.org/ktausch/pylinex}} which was originally developed for fits with linear models \citep{Tauscher:2018a}, but has since been extended to allow for nonlinear models as well.
  
  \subsection{Fit results}
  
\begin{figure}[h!!]
    \centering
    \includegraphics[width=0.48\textwidth]{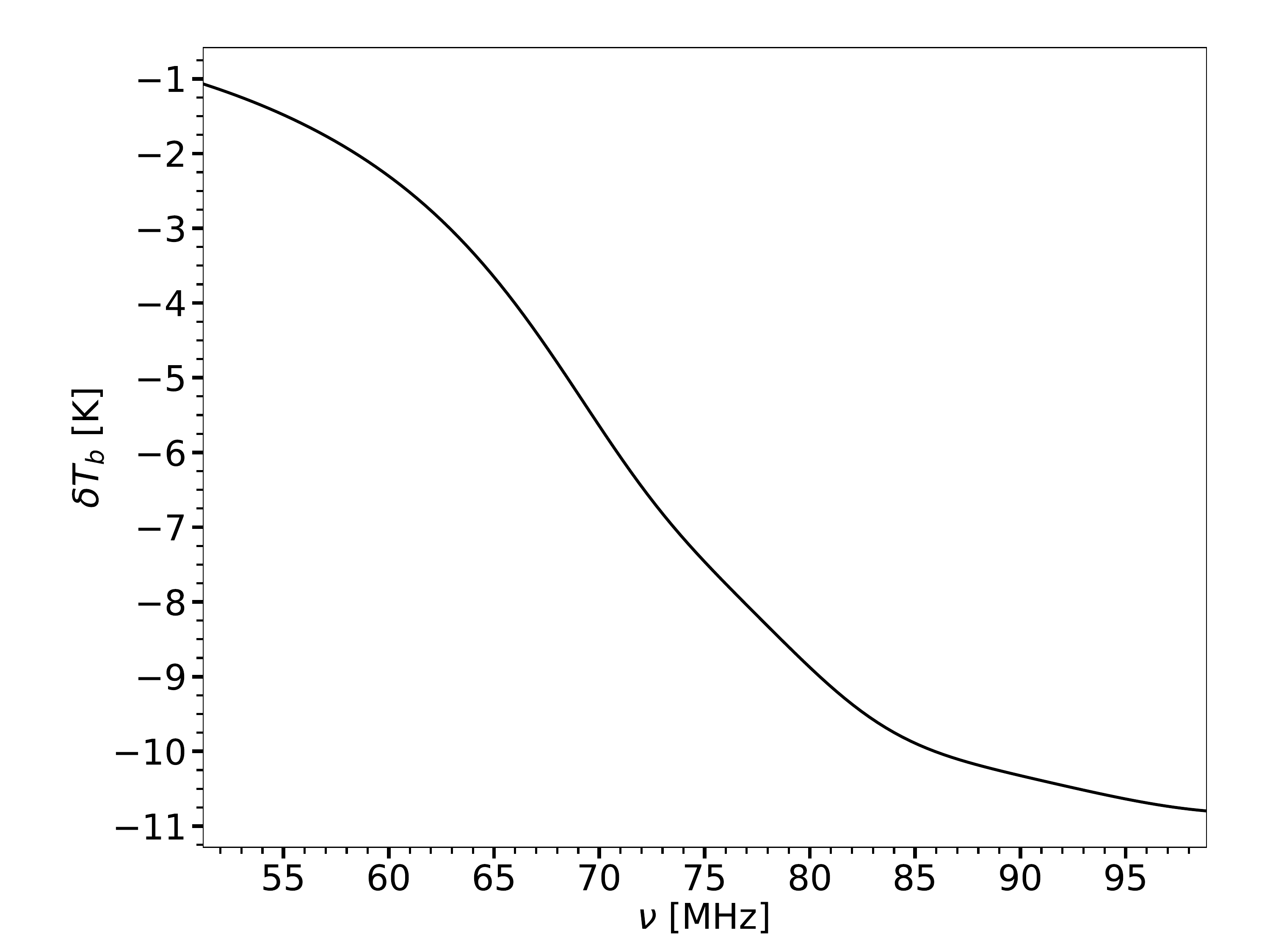}
    \includegraphics[width=0.48\textwidth]{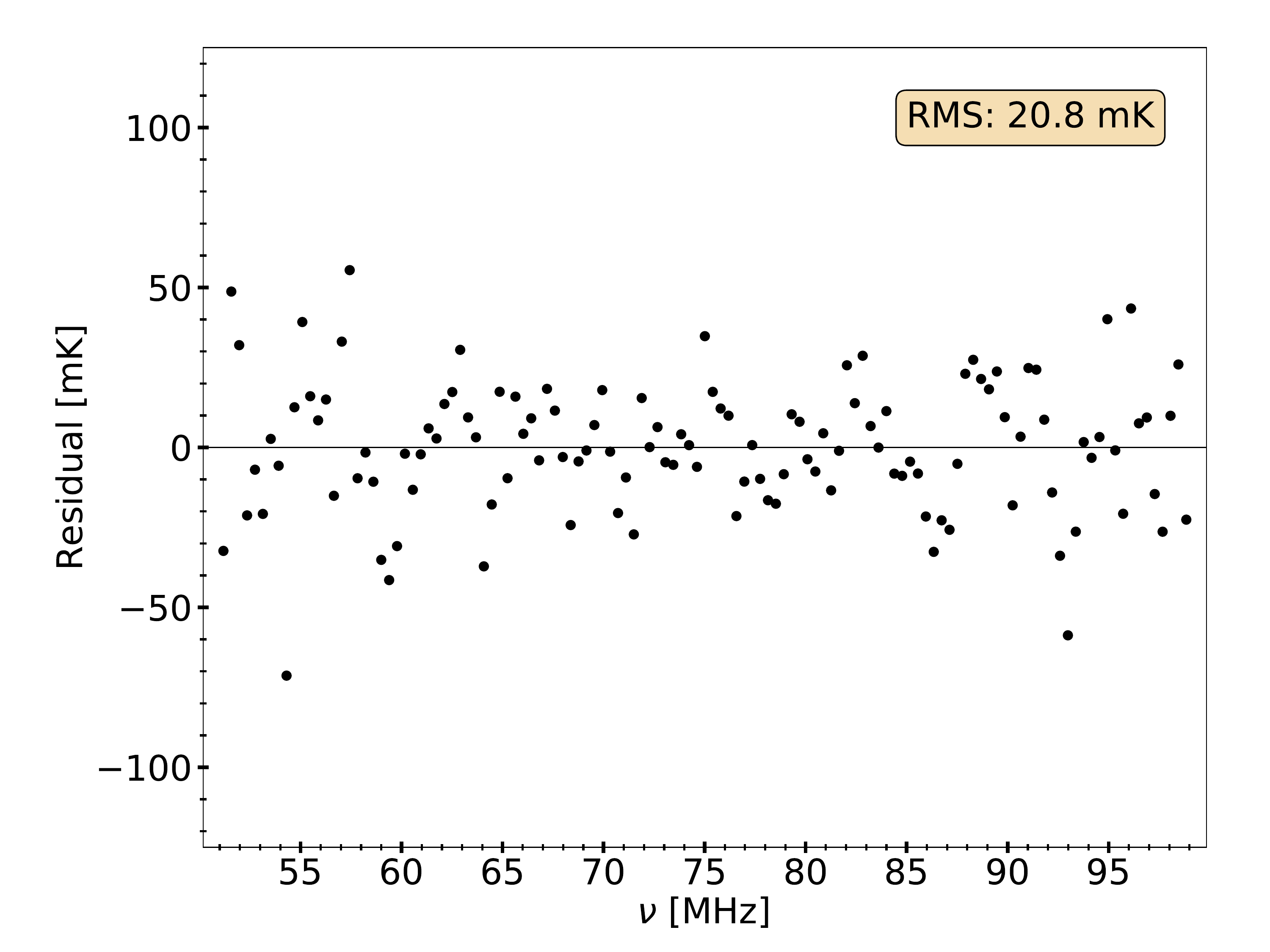}
    \caption{\textit{Left}: The sum of the three fit resonances at 73.8, 84.2, and 111.8 MHz. \textit{Right}: Residuals to the fit which resulted in the resonances in the left panel. The residuals of this fit have an rms of 20.8 mK. The parameters of the three resonances are given in Table~\ref{tab:resonance-fit-from-data}.} \label{fig:fit-to-data-resonances}
  \end{figure}

\begin{figure}[h!!]
    \centering
    \includegraphics[width=0.48\textwidth]{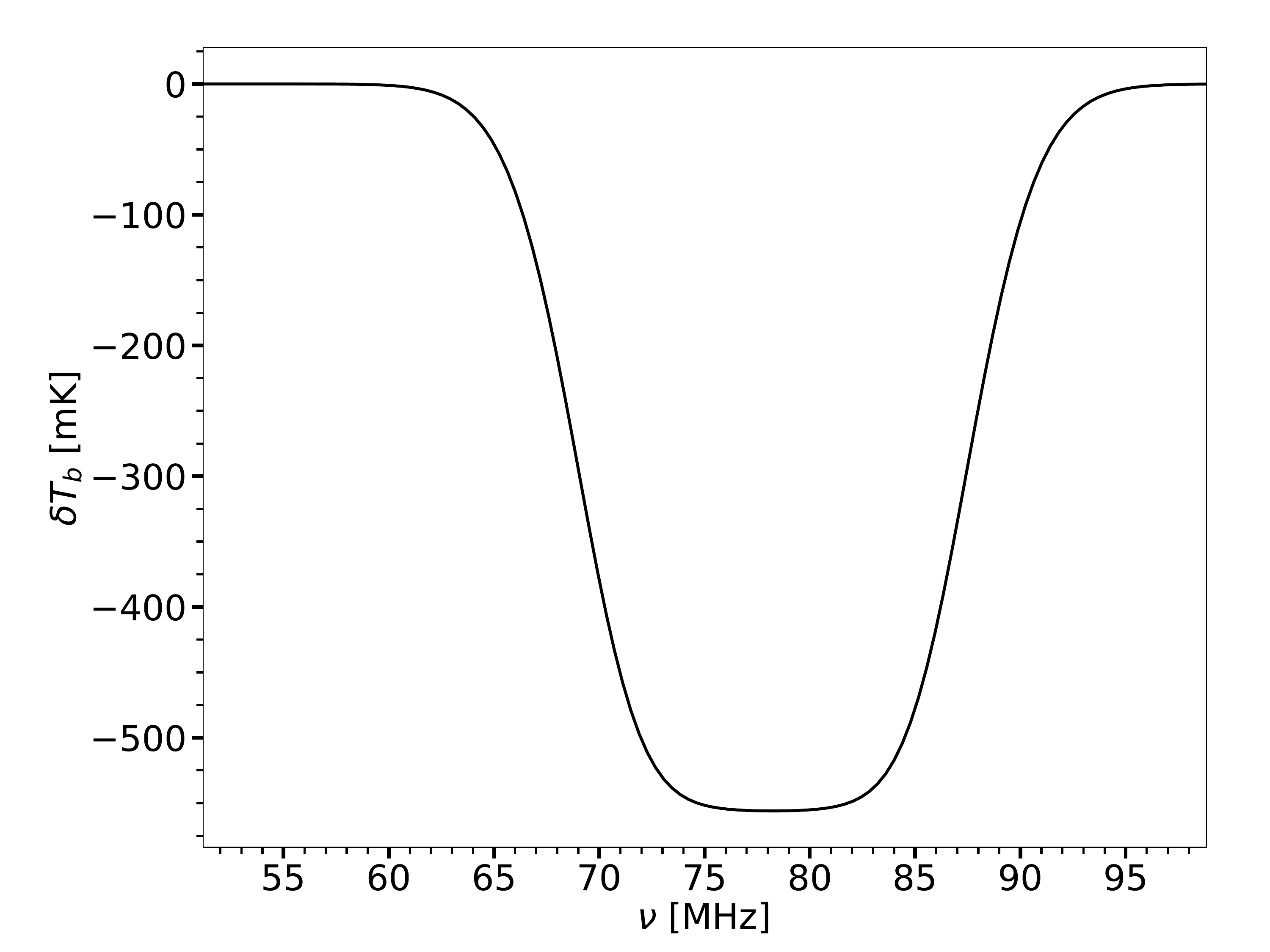}
    \includegraphics[width=0.48\textwidth]{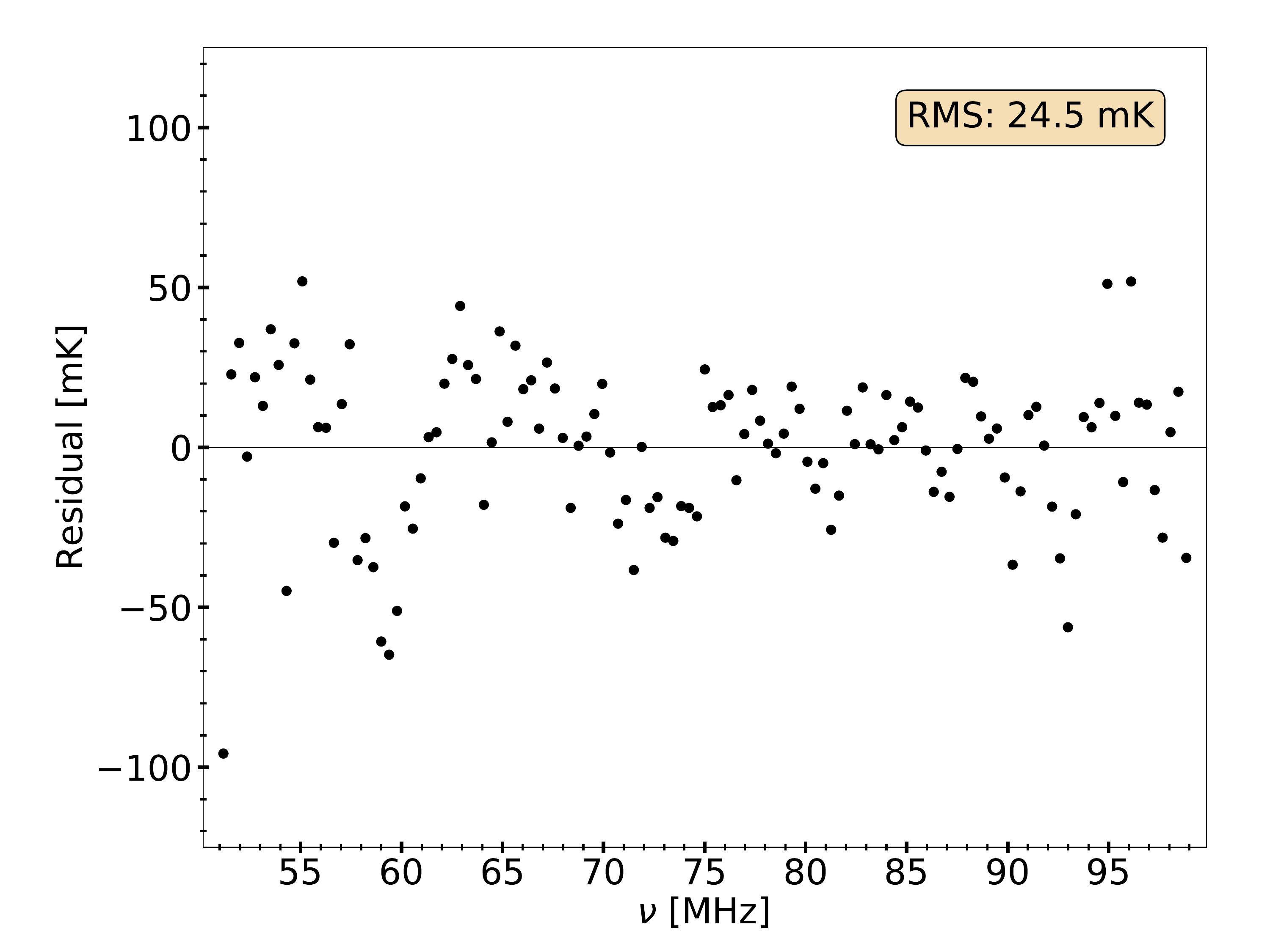}
    \caption{Similar to Fig.~\ref{fig:fit-to-data-resonances}, except the foreground model is given by Eq.~\ref{eq:linearized-physical-foreground-model} and a flattened Gaussian model is used in place of the resonances. This is the same model used in the fit which produced Fig. 1 of \cite{bowman2018absorption}. The residuals of this fit have an rms of 24.5 mK. The flattened Gaussian has a 556 mK amplitude, is centered at 78.2 MHz, has a full width half max of 18.8 MHz, and has a flattening parameter $\tau$ of 5.8.} \label{fig:fit-to-data-flattened-gaussian}
  \end{figure}

  Fig.~\ref{fig:fit-to-data-resonances} summarizes the results of the fit to the data with three resonances and a two-term foreground model. This fit, the parameters of which are shown in Table~\ref{tab:resonance-fit-from-data}, should be compared to the fit in Fig.~\ref{fig:fit-to-data-flattened-gaussian}, which was performed with a flattened Gaussian signal model,
  \begin{equation}
    T_{21} = A\left(\frac{1-e^{-\tau e^B}}{1 - e^{-\tau}}\right) \ \ \text{ where } \ \ B=\frac{4(\nu-\nu_0)^2}{w^2}\ \ln{\left[-\frac{1}{\tau}\ln{\left(\frac{1 + e^{-\tau}}{2}\right)}\right]},
  \end{equation}
  and a foreground model given by
  \begin{equation}
    T_{\text{fg}} = a_0 \nuovernuref^{-2.5} + a_1\nuovernuref^{-2.5}\ln{\nuovernuref} + a_2\nuovernuref^{-2.5}\left[\ln{\nuovernuref}\right]^2 + a_3 \nuovernuref^{-4.5} + a_4 \nuovernuref^{-2}, \label{eq:linearized-physical-foreground-model}
  \end{equation}
  both used for Fig.~1 of \cite{bowman2018absorption}.
  
  \begin{table}[t!!]
    \centering
    \caption{Maximum likelihood parameters for the fitted resonances whose sum is plotted in the left panel of Figure~\ref{fig:fit-to-data-resonances}.}
    \label{tab:resonance-fit-from-data}
    \begin{tabular}{c c c c}
      \hline
      \hline
      Mode & $\nu_0$ & $A(\nu_0)$ & $Q$ \\
      {} & [MHz] & [mK] & {} \\
      \hline
      TM20 & 73.8 & -2235 & 3.9 \\
      TM21 & 84.2 & -2469 & 3.8 \\
      TM30 & 111.8 & -9403 & 1.3 \\
      \hline
    \end{tabular}
  \end{table}
  
  The rms residuals of the resonance and flattened Gaussian fits are $20.8$ mK and $24.5$ mK, respectively, while the two fits have 11 and 9 parameters, respectively. Since the absolute magnitude of the noise level of the data is unknown, the numbers of parameters and rms residuals do not allow us to conclude whether either fit is good in an absolute sense or whether one fit is better than the other; however, their relative values allow us to conclude that the physically motivated resonance model yields another viable explanation of the EDGES data.

\section{Physical Model Parameter Estimation}
\label{sec:Physical}
The dimensions of the patch and properties of the soil may be estimated from the characteristics of the resonances found in Sec.\ref{sec:Fit}.  An analytical model that includes dispersion must be utilized for accuracy. While it is likely that a moisture gradient is present, the details are unknown, so a piecewise homogeneous, layered profile, as depicted in Fig. \ref{fig:EDGESsketch}, was adopted for this proof-of-principle analysis. In addition, a thin dry soil model is invoked ($h\ll\lambda$) where the field variations along the height are considered constant and the electric field is nearly normal to the surface of the patch (except at the edges where fringing occurs).  In this way, the three-dimensional cavity may be treated as a two-dimensional microstrip line resonator.  

The line geometry and boundary conditions restrict the wavenumber, $k$, to discrete values, $k_{mn}$, where $m$ and $n$ are integers.  The constraint can be derived from the homogeneous wave equation involving the vector potential yielding $k_{mn}D=m\pi$ and 
\begin{equation}
\label{eq:lineResonant}
k_{mn}^2 = k_o^2 - \left[\frac{n\pi}{W}\right]^2,
\end{equation}
where $k_o$ is the complex wavenumber or phase constant associated with the dispersive transmission line.  From \cite{collinfoundations}, this is related to the resonant frequencies of the cavity as
\begin{equation}
\label{eq:waveNum}
k_o=2\pi\nu_{mn}\sqrt{\mu\epsilon}.
\end{equation}

The determination of $k_o$ involves the soil properties.  The electrical conductivities of the soil, given in Table \ref{tab:soil}, are quite small and will greatly limit the Q of the resonances. However, the permittivity of the dry soil is a complex quantity, $\epsilon=\epsilon'-i\left( \epsilon''+ \frac{\sigma_d}{2\pi\nu} \right)$, affecting the propagation constant.  If magnetizable minerals are present in the soil, the permeability may also be complex, with $\mu=\mu'-i\mu''$ to include damping.  The moist soil, a lossy ohmic conductor, adds a frequency dependent internal inductance, $L_{int}$, to the distributed model due to the skin effect which also affects the propagation constant. This is described as 
\begin{equation}
\label{eq:LI}
L_{int} = \frac{1}{2} \sqrt{\frac{\mu_o}{\pi\sigma_w\nu_{mn}}},
\end{equation}
where $\sigma_w$ is the conductivity of the moist soil.  

As a result, the propagation constant is the imaginary part of $ik_o$ given by
\begin{equation}
\label{eq:beta}
Im[ik_o]=2\pi\nu_{mn} \sqrt{\frac{\epsilon'\mu'-\left( \epsilon''+ \frac{\sigma_d}{2\pi\nu_{mn}} \right)\mu''}{2}\left[\sqrt{1+\left(\frac{\epsilon'\mu''+\left( \epsilon''+ \frac{\sigma_d}{2\pi\nu_{mn}} \right)\mu'}{\epsilon'\mu'-\left( \epsilon''+ \frac{\sigma_d}{2\pi\nu_{mn}} \right)\mu''}\right)^2} +1 \right]}=2\pi\nu_{mn}\sqrt{L_LC_L}.
\end{equation}

A TEM transmission line formed by the metal patch, thin dry soil layer ($h\ll\lambda$ assumed), and underlying conductor can be described in a unique manner by a distributed-parameter electric network consisting of series inductance ($L_L$) and shunt capacitance ($C_L$) per unit length for energy storage and series resistance ($R_L$) and shunt conductance ($G_L$) per unit length for energy dissipation.  

Under the conditions of small dielectric loss ($\epsilon''\approx 0$) and no magnetic loss ($\mu''\approx 0$), Eq.~\ref{eq:beta} becomes separable yielding
\begin{equation}
\label{eq:CL}
C_L=\frac{\epsilon'}{2} \left[\sqrt{1+\left( \frac{\sigma_d} {2\pi\nu_{mn}\epsilon'} \right)^2} +1 \right]
\end{equation}
and
\begin{equation}
\label{eq:LL}
L_L=L_{int}+L_{ext} = L_{int}+\left(\frac{Z_{co}}{S}\right)^2 C_L = \mu'
\end{equation}
where, from \cite{pozar1990microwave}, the characteristic impedance of the lossless microstrip quasi-TEM line is given by
\begin{equation}
\label{eq:Zline}
Z_{co} = \sqrt{\frac{L_{ext}}{C_L}} = \frac{120\pi}{\sqrt{\epsilon_e}\left[W/h+1.393+0.677\ln\left(W/h +1.444\right)  \right]} = \frac{120\pi S}{\sqrt{\epsilon_e}}
\end{equation}
with $S$ as the dimensionless geometrical factor relating the line impedance to the characteristic impedance of the medium. $L_{ext}$ is the inductance per unit length of the line when the conductors are lossless.   

Finally, the $Q$ of a resonance is the ratio of time-average energy stored to energy dissipated per unit length at the given resonant frequency.  Since the loss is dominated by the lower conductor (i.e. for the dry soil, $\alpha = Re[ik_o] \approx 0$ yielding $G_L \approx 0$), then $Q_{mn} \approx2\pi\nu_{mn} L_{ext} / (R_L S)$ with
\begin{equation}
\label{eq:RI}
R_L = \frac{1}{2}\sqrt{\frac{\pi\mu_o\nu_{mn}}{\sigma_w}},
\end{equation}
which is derived from the skin depth analysis.  Since the $Q$ is defined on a per unit length basis, it must be scaled by the number of half wavelengths present at the given resonance.  

These equations were used to estimate the frequencies and Q's of the three in-band resonances and the band-adjacent resonances, given physically realizable values for $D$, $W$, $h$, $\epsilon'$, $\sigma_d$, and $\sigma_w$. The results are summarized in Table \ref{tab:parameter_estimates}.  It should be stressed that this solution is not unique but represents one plausible configuration that fits the available information. As expected, from Table \ref{tab:resonances}, TM22 is nearly degenerate with TM30 and cannot be resolved given the low $Q$ factors.   

  \begin{table}[h!]
    \centering
    \caption{Estimated physical parameters of the patch absorber.}
    \label{tab:parameter_estimates}
    \begin{tabular}{c c c c c c c c c c c c c c c c}
      \hline
      \hline
      W & D & h & $\epsilon_r'$ & $\sigma_d$ & $\sigma_w$ & $Q_{11}$ & $Q_{20}$ & $Q_{21}$ & $Q_{22}$ & $Q_{30}$ & $\nu_{11}$ & $\nu_{20}$ & $\nu_{21}$ & $\nu_{22}$ & $\nu_{30}$ \\
      $[m]$ & $[m]$ & $[m]$ & & $[S/m]$ & $[S/m]$ & & & & & & $[MHz]$ & $[MHz]$ & $[MHz]$ & $[MHz]$ & $[MHz]$\\ 
      \hline 1.9 & 2.1 & 0.2 & 4.0 & 0.0005 & 1.0 & 3.4 & 2.8 & 2.6 & 2.4 & 2.3 & 55.0 & 73.8 & 84.2 & 110.7 & 111.0\\
      \hline
    \end{tabular}
  \end{table}

The characteristics of the two dominant resonances in Table \ref{tab:parameter_estimates}, TM20 and TM21, were anchored to the fit results of Table~\ref{tab:resonance-fit-from-data}.  However, the analytical model suggests that the frequencies of the TM22/TM30 modes are approximately 0.7 $\%$ lower.  The more weakly coupled TM11 mode shown here is likely below the residual noise in the data.  

\section{Experimental and simulation tests}
\label{sec:Consider}
\subsection{Moist Soil}
A rudimentary experiment was conducted at the Green Bank Observatory to confirm that nominal moisture content of soil is sufficient to form the lower conductor of the patch antenna. From \cite{eCFR2018}\footnote{\url{https://www.gpo.gov/fdsys/pkg/CFR-2004-title47-vol4/pdf/CFR-2004-title47-vol4-sec73-190.pdf}}, the estimated soil conductivity for the Green Bank, WV region is 0.002 $S/m$.  In contrast with the soil at the EDGES site, the moist soil here extends to the surface, with the air and pine wood frame under the mesh serving as the dielectric. The patch was made from 3 $m$ x 3 $m$, \#23 gauge, galvanized, welded steel mesh (4 squares per inch) attached to a pine frame with double-folded, overlapping seams.  

The patch is an unbalanced antenna that can be fed by way of a coaxial cable.  For the ground connection, an aluminum rod was driven approximately 50 $cm$ into the soil adjacent to the edge of the patch.  The outer shield of a coaxial cable was clamped to the rod and the inner conductor was attached to the mesh using a spring clip. The height of the mesh above ground was set using concrete blocks positioned at the corners while the entire patch was moved laterally to change the feed point position.  

\begin{figure}[h!!]
    \centering
    \includegraphics[width=0.42\textwidth]{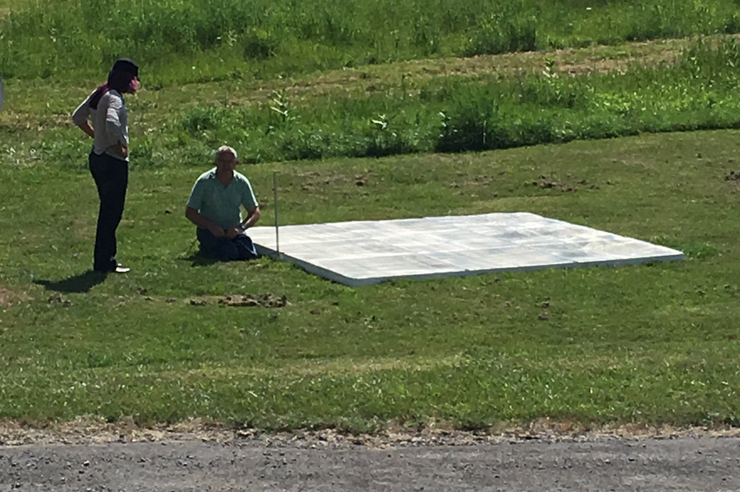}
    \includegraphics[width=0.57\textwidth]{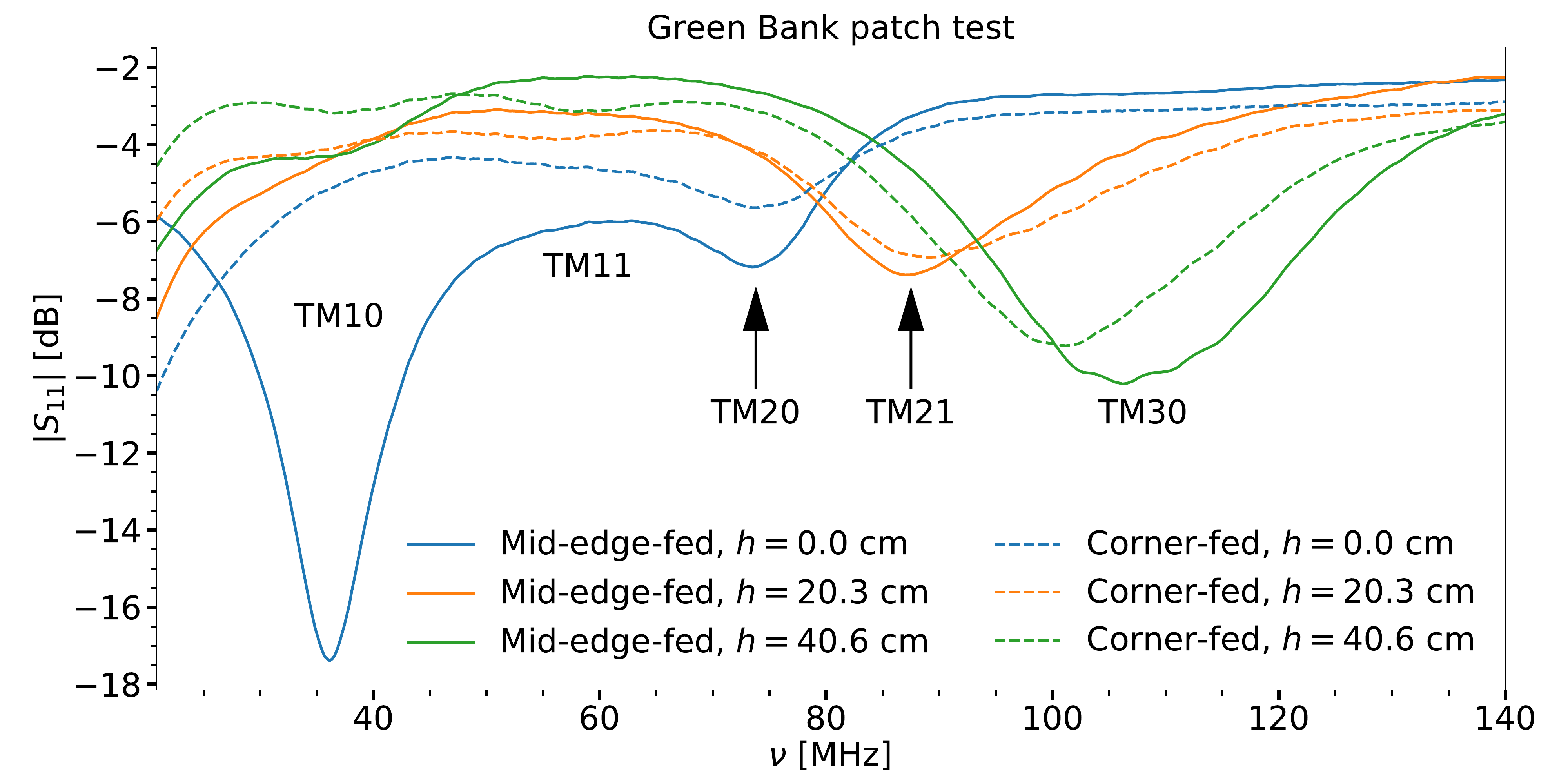}
    \caption{\textit{Left}: Photograph of the patch antenna in Green Bank, WV used to confirm TM-mode resonances for the case when moist ground  ($\sigma_w=0.002$ $S/m$) acts as the lower conductor. The Mid-Edge feed point is illustrated.    
\textit{Right}: Spectrum of reflected power from the patch for several heights of the patch above ground and feed point locations. Resonant modes are indicated.}
\label{fig:GBresonances}
  \end{figure}

The patch was excited by an Anritsu Model MS2024A/15 Vector Network Analyzer (VNA) attached to the other end of the coaxial cable (characteristic impedance $Zo=50$ $\Omega$) located approximately 30 $m$ from the patch.  One-port network calibration was performed at the antenna end of the cable using the coaxial open, short, and 50 $\Omega$ load termination standards provided by the manufacturer of the VNA.  The plot shown in Fig.~\ref{fig:GBresonances} is referenced to this calibration plane and shows the TM-mode resonances that occurred in the patch for several heights of the mesh above ground and for two feed locations - one at the midpoint of the patch edge and the other at the corner.  The results of this experiment confirm that soil having an electrical conductivity as low as 0.002 S/m will form a lossy, lower conductor for the patch antenna. From Table \ref{tab:soil}, lower soil moisture content at the MRO site can yield conductivities several orders-of-magnitude greater than this value.  

\subsection{Beam Patterns of the Patch Absorber}
\label{subsec:beams}
The commercial electromagnetic simulation package Microwave Studio by CST was used to estimate the beam patterns of the patch absorber. Fringing electromagnetic fields across a gap in the ground plane allow a fraction of the energy carried by the impinging plane waves, normally confined to the upper half-space, to excite the patch structure within the soil.

The model used in this simulation was a 20 $m$ square ground plane with a 2 $m$ square patch located at the center.  A 5 $cm$ gap was included around the patch for fringe field coupling.  The soil structure under the ground plane consisted of 20 $cm$ thick dry component ($\epsilon_r=4.0$, and $\sigma_d=0.0005$ $S/m$) with moist soil ($\sigma_w = 1.0$ S/m) underneath.  For beam pattern estimation, the patch was excited by a discrete port located midway along the edge of the patch and extending in the z-direction downward from the patch to the moist soil.  

\begin{figure}[h!!]
    \centering
    \includegraphics[width=0.48\textwidth]{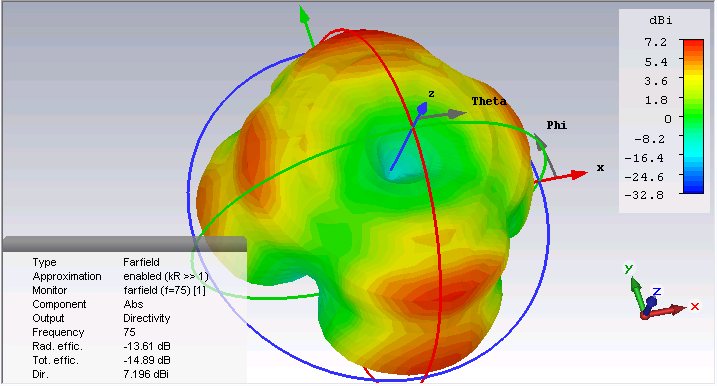}
    \includegraphics[width=0.48\textwidth]{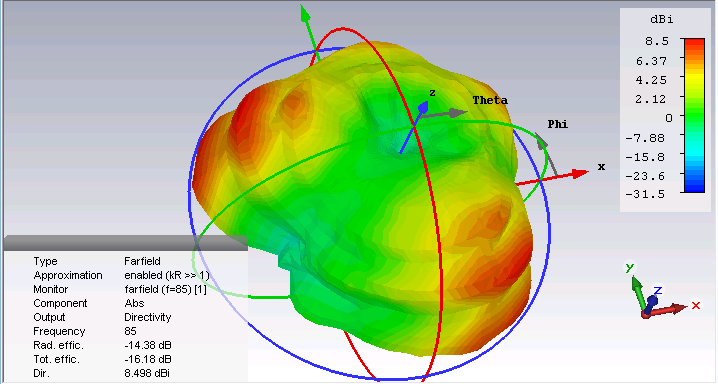}
    \caption {Representative plots of the patch absorber's beam patterns: \textit{Left}: at 75 MHz, and \textit{Right}: at 85 MHz.}
\label{fig:beams}
  \end{figure}

Plots of the simulated beam patterns are given in Fig. \ref{fig:beams}.  The most outstanding feature of the beams is the multi-lobed response far removed from the bore sight of the dipole antenna (+z axis).  The left panel shows the pattern at 75 MHz where the TM20 mode dominates while the right panel shows the rotated pattern at 85 MHz where the TM21 mode becomes engaged.

There are two important points regarding these patterns.  First, the resonances, TM20 and TM21, are each excited by different regions of the sky. Second, the amplitudes of the absorption features should each be time-varying as the sources traverse through the complicated beam patterns. Hence, the broad spectral feature that is formed by these two resonances closely spaced in frequency will each change characteristics independently as a function of time.   

\subsection{Plane Wave Excitation}
\begin{figure}[h!!]
    \centering
    \includegraphics[width=0.48\textwidth]{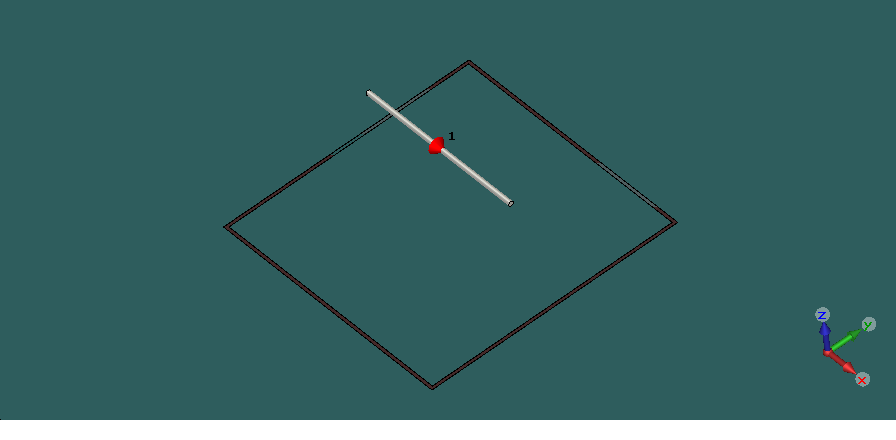}~~
   \includegraphics[width=0.49\textwidth]{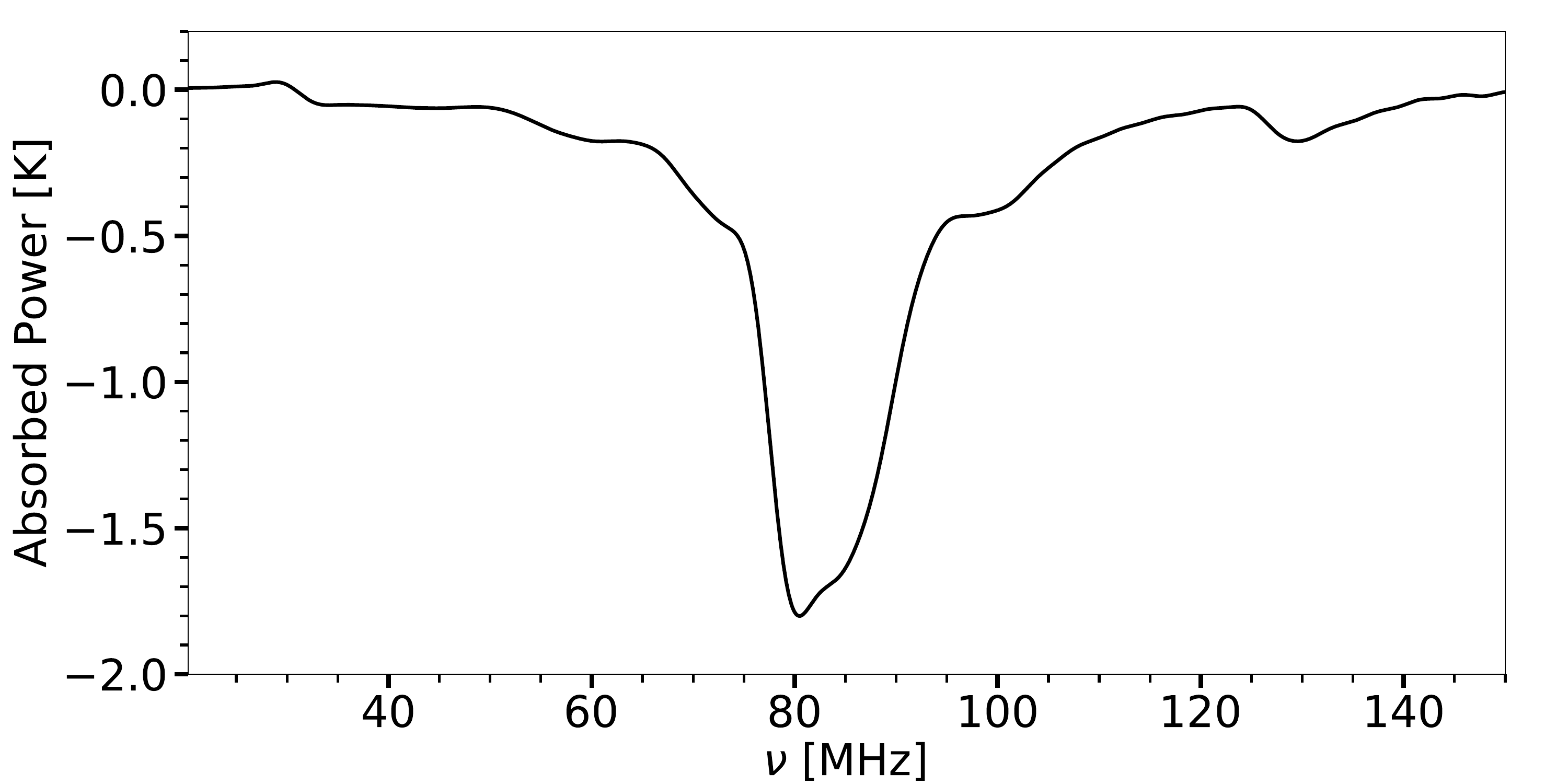}
\caption{\textit{Left}: A rendering from the simulator model showing the dipole above the 2 $m$ x 2 $m$ patch at the center of the ground plane. \textit{Right}: Spectral plot of missing power at the x-axis oriented dipole terminals due to the presence of the patch for the single plane wave excitation (x-oriented polarization) from a 4,000 $K$ sky region.}
\label{fig:planeWave}
\end{figure}

To gauge the absorption characteristics of the patch, a simple dipole antenna was added to the simulation model described in Sec.~\ref{subsec:beams}.  It was tuned for 91 MHz and placed 80 cm above the ground plane, as shown in the left panel of Fig.~\ref{fig:planeWave}. A plane wave, with E-field orientation in the $x$ direction, propagates in the $-y$ direction toward the dipole from an elevation of 45 degrees above the ground plane. The right panel of Fig.~\ref{fig:planeWave} shows the power loss in the ground plane compared to that of a solid ground plane, assuming the plane wave emanates from a (nominal) 4,000 $K$ sky region.  The primary in-band absorption resonances are clearly visible with magnitudes comparable to the fit resonances.  This simulation illustrates that a patch antenna embedded in the ground plane can imprint spectral structure on the power received by the dipole.  The magnitudes of such features can be quite difficult to discern from dipole reflection coefficient and beam pattern measurements since a dynamic range of greater than 30 $dB$ is required.  

\section{Conclusions}
\label{sec:Conclusion}
A ground screen artifact that can produce  absorption features in the spectrum of Global 21-cm experiments was described.  A plausible physical model, based on patch antenna theory and defined by physical parameters, was developed and the associated in-band and adjacent band resonances of the patch antenna cavity modes were compared with those found from fitting to the published EDGES data.  The ground screen absorbing characteristics were discussed along with examples of patch antenna beam patterns. 

The conclusions may be summarized as follows:
\begin{itemize}
\item The three cavity mode resonant features of the ground plane patch absorber together with an $N=2$ term polynominal foreground model fits the published EDGES data (11 parameters, 20.8 mK rms) as well as that of the flattened-Gaussian feature model with a $5$ term foreground model (9 parameters, 24.5 mK rms). The significance of the difference between the numbers of parameters and rms residual values of the two fits can only be determined from a goodness-of-fit statistic calibrated to the noise level.
\item The parameters defining the geometrical and soil characteristics estimated from the fitting results that used an analytical model of the patch absorber are consistent with the physical environment of the instrument and the spectral trough in the data.
\item Measurements conducted at the Green Bank Observatory illustrate that soil moisture can form the lower conductor for the resonant patch absorber.  However, mineral deposits may also be a factor.  
\item Electromagnetic simulations show that an absorption trough can be imprinted onto the sky spectrum acquired by a dipole when it is located above a ground plane containing an isolated patch.   
\item High order polynomial foreground models are dangerous for two main reasons: a) they can fit out signal or hidden systematic features that are as spectrally smooth as the foreground, and b) it is unclear whether they can model foreground emission to precision levels necessary for accurate extraction of the 21-cm global signal.
\item Beam chromaticity and instrument systematics are serious problems, especially when they interact with generic flexible models like polynomials.
\end{itemize}

\acknowledgments
We wish to thank J. Bowman, A. Rogers, and R. Monsalve for their very helpful discussions of the EDGES instrument and data analysis.  We acknowledge the contributions of J. Hewitt, B. Nhan, L. Aguirre, J. Bracks, and D. Riley during our ground screen measurements at the Green Bank Observatory.   
This work was directly supported by the NASA Solar System Exploration Virtual Institute cooperative agreement 80ARCC017M0006. DR is supported by a NASA Postdoctoral Program Senior Fellowship at the NASA Ames Research Center, administered by the Universities Space Research Association under contract with NASA.  The National Radio Astronomy Observatory is a facility of the National Science Foundation operated under cooperative agreement by Associated Universities, Inc.

%

\vspace{5mm}
\facilities{National Radio Astronomy Observatory, Green Bank Observatory}


\software{CST Microwave Studio, Numpy, Scipy, and Matplotlib Python packages}     



\bibliography{PatchAbsorber}






\end{document}